\documentclass[conference, a4paper]{IEEEtran}

\usepackage[font=footnotesize,caption=false]{subfig} 

\usepackage{cite}
\usepackage{amsmath,amssymb,amsfonts}
\usepackage{algorithmic}
\usepackage{graphicx}
\usepackage{textcomp}
\usepackage{xcolor}
\usepackage{booktabs}
\usepackage{glossaries}
\usepackage{subfig}
\usepackage{algorithm}
\usepackage{algorithmic}
\usepackage{bbm}
\usepackage{multirow}
\newcommand{\set}{\mathcal}

\newacronym{leo}{LEO}{Low-Earth Orbit}
\newacronym{isl}{ISL}{Inter-Satellite Link}

\DeclareMathOperator{\sgn}{sgn}
\DeclareMathOperator*{\maximize}{maximize}

\DeclareMathOperator*{\argmax}{arg\,max}

\def\BibTeX{{\rm B\kern-.05em{\sc i\kern-.025em b}\kern-.08em
    T\kern-.1667em\lower.7ex\hbox{E}\kern-.125emX}}

\usepackage{etoolbox}

\begin{document}

\title{\vspace{0.03in}Inter-Plane Inter-Satellite Connectivity in LEO Constellations: Beam Switching vs.\ Beam Steering}
\author{\IEEEauthorblockN{Israel Leyva-Mayorga\IEEEauthorrefmark{1}, Maik R{\"o}per\IEEEauthorrefmark{2}, Bho Matthiesen\IEEEauthorrefmark{2}\IEEEauthorrefmark{3}, Armin Dekorsy\IEEEauthorrefmark{2}, Petar Popovski\IEEEauthorrefmark{1}\IEEEauthorrefmark{3}, and Beatriz Soret\IEEEauthorrefmark{1}}

\IEEEauthorblockA{\IEEEauthorrefmark{1}Department of Electronic Systems, Aalborg University, Aalborg, Denmark}
\IEEEauthorblockA{\IEEEauthorrefmark{2}Gauss-Olbers Center, c/o University of Bremen, Dept. of Communications Engineering, Germany}
\IEEEauthorblockA{\IEEEauthorrefmark{3}University of Bremen, U Bremen Excellence Chair, Dept.\ of Communications Engineering, Germany}
\IEEEauthorblockA{Email: \{ilm, petarp, bsa\}@es.aau.dk, \{roeper, matthiesen, dekorsy\}@ant.uni-bremen.de}
}

\maketitle

\begin{abstract}
    Low Earth orbit (LEO) satellite constellations rely on inter-satellite links (ISLs) to provide global connectivity. However, one significant challenge is to establish and  maintain \emph{inter-plane ISLs}, which support communication between different orbital planes. This is due to the fast movement of the infrastructure and to the limited computation and communication capabilities on the satellites.
    In this paper, we make use of antenna arrays with either Butler matrix beam switching networks or digital beam steering to establish the inter-plane ISLs in a LEO satellite constellation. Furthermore, we present a greedy matching algorithm to establish inter-plane ISLs with the objective of maximizing the sum of rates. This is achieved by sequentially selecting the pairs, switching or pointing the beams and, finally, setting the data rates. Our results show that, by selecting an update period of 30~seconds for the matching, reliable communication can be achieved throughout the constellation, where the impact of interference in the rates is less than 0.7\,\% when compared to orthogonal links, even for relatively small antenna arrays. Furthermore, doubling the number of antenna elements increases the rates by around one order of magnitude.
\end{abstract}

\section{Introduction}
\label{sec:intro}

In \gls{leo} satellite constellations, where satellites are organized in several orbital planes, implementing direct inter-satellite communication through inter-satellite links (ISLs) is the sensible choice to provide global service to delay-sensitive applications~\cite{Leyva-Mayorga2021}. 
Achieving efficient inter-satellite communication is challenging since the constellation is a moving infrastructure with satellites orbiting the Earth at around $7.5$\,km/s.
While the intra-plane ISLs, connecting satellites from the same orbital plane, are rather stable, the inter-plane ISLs, connecting satellites in different orbital planes, are greatly dynamic. Hence, connection times to potential neighbors vary widely, even in fully symmetric constellations~\cite{Akyildiz2019}. Furthermore, the risk of collisions between satellites is minimized by orbital separation, i.e., deploying the satellites in orbital planes at slightly different altitudes~\cite{Lewis2019}. Nevertheless, orbital separation leads to asymmetry in the orbital periods and to frequent changes in the relative positions of the satellites, further challenging the adaptation of the inter-plane ISLs.

The selection of a proper antenna technology is essential to achieve efficient inter-plane communication. Free-space optical (FSO) technologies can provide high data rates, ultra-narrow beams, and reduced antenna size. However, the main challenge to achieve inter-plane communication with FSO is correctly pointing the antennas. This can be avoided by resorting to more traditional wireless technologies. Unlike the S- and C-frequency bands, employing the K- and Ka-bands also enables the use of large bandwidths and sufficiently small antenna elements to make antenna arrays feasible even for small satellites. This offers great design flexibility in terms of beamwidth, gain, and beam pointing technology.

With the advent of millimeter-wave and 5G, Butler matrix beamforming networks have gained relevance in terrestrial communications~\cite{Chang2010,Yu2018}. These are cost-efficient and low-complexity beam switching networks that produce a series beams in pre-defined directions~\cite{Zooghby2005, Wang2018}. These beams are switched by simply feeding one or more of the input ports, which offers an interesting trade-off between performance, cost, and complexity of operation and implementation. In comparison, digital beam steering is able to precisely point the beams in the desired direction and, hence, is greatly attractive to combat the fast orbiting velocities in LEO satellite communications~\cite{Su2019}. Nevertheless, beam steering is much more complex than switching, requiring variable phase shifters to manipulate the input signals in each antenna element.

In our previous work~\cite{Leyva-Mayorga2021}, 
we studied the connectivity potential of the inter-plane ISL, providing algorithms for the dynamic establishment of these links in LEO constellations with the objective of maximizing the sum of rates. We considered the extreme cases of satellites with either isotropic antennas or ultra-narrow beam antennas with perfect pointing capabilities. 
Once the ISLs are established, the transport capacity can be calculated~\cite{Jiang2020}. However, this requires to define source-destination pairs. On the other hand, the number of ISLs in a constellation has been used as a connectivity metric to design LEO constellations~\cite{Kak2019} and the sum of rates in the ISLs can be seen of an extension of such metric. Hence, in this paper, we focus our attention on the establishment of the inter-plane ISLs with realistic models of antenna arrays fed by either a Butler matrix or by a digital beamformer with beam steering. This introduces an additional level of complexity to the matching problem, which has to pair now not simply satellites, but to switch or steer the beams to maximize the sum of rates.


\section{System model}
\label{sec:system_model}
We consider the inter-satellite transceiver matching problem in a general LEO constellation where $N$ satellites are evenly distributed in $P$ circular and evenly-spaced orbital planes. Each orbital plane $p\in\{1,2,\dotsc,P\}$ is deployed at a given altitude above the Earth's surface $h_p$~km, at a given longitude $\epsilon_p$~radians, at a given inclination $\delta$, and consists of $N_p$ evenly-spaced satellites. For notation simplicity, we define the function $p(\cdot)$ to be the orbital plane in which a satellite is deployed. 
We set an orbital separation $\Delta h$, which determines the difference in the altitude of contiguous orbital planes~\cite{Lewis2019}. Building on this, the altitude of a given orbital plane $p>1$ is $h_p=h_{p-1}+\Delta h$ and orbital plane $p=1$ is deployed at the minimum altitude $h_1$. 

We model the constellation at any given time instant $t$ as a weighted undirected graph \mbox{$\set{G}_t=(\set{V},\set{E}_t)$} where $\set{V}$ is the set of vertices (satellites) and $\set{E}_t$ is the set of undirected edges (feasible inter-plane ISLs) at time $t$. Graph $\set{G}$ is multi-partite with $P$ vertex classes $\set{V}_1,\set{V}_2,\dotsc,\set{V}_P$. We denote an undirected edge as $uv$ and a source-destination pair as $(u,v)$, where $u,v\in\set{V}:p(u)\neq p(v)$. The weight of an edge $uv$ at time $t$ is denoted as $w_t(uv)=w_t(vu)$. 

Let $\mathbf{r}^{(u)}_t=\left[x_t^{(u)},y_t^{(u)},z_t^{(u)}\right]^\intercal$ be the column vector with the cartesian coordinates of satellite $u$ at time $t$. From there, Euclidean distance between two satellites, denoted as $l(u,v,t)$, can be easily calculated. Inter-satellite communication is affected by free-space path loss (FSPL), by thermal noise -- which is additive white Gaussian (AWGN)~\cite{3GPPTR38.821} --, and by the interference from other satellites. In addition, the Earth blocks the line of sight (LoS) between two satellites $u$ and $v$ in orbital  if $l(u,v,t)>l^*(p,q)$, where $l^*(p,q)$ is the maximum slant range (i.e., line-of-sight distance) between two satellites in orbital planes $p=p(u)$ and $q=p(v)$. Assuming the Earth is a perfect sphere we have 
\begin{IEEEeqnarray}{c}
    l^*(p,q)= \sqrt{h_{p}^2+2\mathrm{R_E}h_p}
    +\sqrt{h_{q}^2+2\mathrm{R_E}h_q},
    \label{eq:max_l_LoS}
\end{IEEEeqnarray}
where $\mathrm{R_E}$ is the radius of the Earth. 

Hence, the set of edges with no line of sight (NLoS) at time $t$ is 
$\left\{uv\in\set{E}_t:l(u,v,t)>l^\star(p,q)\right\}$. Building on this, the FSPL between $u$ and $v$ when $l(u,v,t) \leq l^\star(p,q)$ is given by $L(u,v,t)=\left(\frac{4\pi l(u,v,t) f}{\mathrm{c}} \right)^2$
and is infinity otherwise. Here $f$ is the carrier frequency and $\mathrm{c}$ is the speed of light.

Each satellite is equipped with two transceivers for unicast inter-plane inter-satellite communication, which allows to maintain up to one active ISL with both neighboring orbital planes. Hence, each satellite can maintain up to one inter-plane ISL at each side of the pitch axis, namely, in direction $d_a\in\{-1,1\}$. The antennas used for inter-plane communication are located at each side of the pitch axis of the satellites. We assume that the antennas and wireless resources for intra-plane communication are independent and do not cause interference to the inter-plane ISLs and vice versa.

To calculate the antenna gains, we define \mbox{$\mathbf{r}^{(u,v)}_t = \left[x_t^{(u,v)},y_t^{(u,v)},z_t^{(uv)}\right]$} as the relative position of satellite $v$ w.r.t. $u$ at time $t$, 
where 
$x_t^{(u,v)}$ and $y_t^{(u,v)}$ denote the position of $v$ w.r.t. $u$'s pitch and roll axis, respectively.

Furthermore, we can define the relative direction of $v$ from $u$ at any time $t$ in terms of $\phi_t^{(u,v)}$ and $\Theta_t^{(u,v)}\in\left[0,\pi\right]$, the azimuth and the polar angles, respectively.
Building on this, we define set of satellites that are located in the direction of antenna $d_a$ w.r.t. the pitch axis of satellite $u$ at time $t$ as $\mathcal{V}_u(d_a, t) =\left\{v\in\mathcal{V}:\cos\left(\phi_t^{(uv)}\right)d_a>0\right\}$. That is, the antenna that must be used for communication from $u$ to $v$ at time $t$ is simply given by $\cos\left(\phi_t^{(uv)}\right)$. 




In order to provide a high antenna gain with a low implementation complexity, we consider a planar $K\times K$ antenna array fed by a Butler matrix with $K\in2^\mathbb{N}$ ports~\cite{Zooghby2005}. These antenna ports can be fed individually to produce $K$ different and orthogonal beams $\mathbf{b}_{k}\in\mathbb{C}^{K^2\times 1}$, for $k=\{1,2,...,K\}$ along the azimuth plane~\cite{Zooghby2005} -- along the angle $\phi$ --, whose elevation $\theta$ is fixed. Since there are two antenna arrays and their corresponding Butler matrices per satellite in directions $d_a\in\{-1,1\}$, a total of $2K$ different beams can be produced at each satellite. Therefore, we denote a specific antenna port at satellite $u$ as $k_a^{(u)}\in\{d_a,2d_a,\dotsc, Kd_a\}$. The radiation pattern of the beams is defined by the number of antenna elements $K\times K$, the distance between them $d_e$, the wavelength $\lambda=c/f$, and the fixed elevation $\theta$. For instance, increasing $K$ increases the number of beams and decreases the beamwidth, which in turn increases the maximum gain. 

Let $G(u,v,k_a^{(u)}, k_a^{(v)}, t)$ be the antenna gain between satellites $u$ and $v$ at time $t$, with antenna ports $k_a^{(u)}$ and $k_a^{(v)}$ at satellites $u$ and $v$, respectively. 

This allows us to calculate the $K$-dimensional steering vectors from an array in satellite $u$ to satellite $v$ at $t$ for the corresponding azimuth angle as
\begin{IEEEeqnarray*}{c}
  \mathbf{a}^{(u,v)}_{t,\text{az}}=\left[1, e^{\frac{-j2\pi d_e}{\lambda}\sin\left(\phi_t^{(u,v)}\right)},\dotsc, e^{\frac{-j2\pi d_e(K-1)}{\lambda}\sin\left(\phi_t^{(u,v)}\right)}\right]^\intercal\IEEEnonumber \\
\end{IEEEeqnarray*}
and for the polar angle as
\begin{IEEEeqnarray*}{c}
 \mathbf{a}^{(u,v)}_{t,\text{pol}}\!=\!\left[1, e^{\frac{-j 2\pi d_e}{\lambda}\cos\left(\Theta_t^{(u,v)}\right)}\!,\dotsc , e^{\frac{-j2\pi d_e(K-1)}{\lambda}\cos\left(\Theta_t^{(u,v)}\right)}\right]^\intercal\!\!. \IEEEnonumber\\
\end{IEEEeqnarray*}
 
 Then, the overall steering vector $\mathbf{a}^{(u,v)}\in\mathbb{C}^{K^2\times 1}$ is given by their Kronecker product
$
    \mathbf{a}_t^{(u,v)} = \mathbf{a}^{(u,v)}_{t,\text{pol}} \otimes \mathbf{a}^{(u,v)}_{t,\text{az}}\,.
$

To calculate the antenna gain with Butler matrix beamforming, let $\mathbf{b}_{k,\text{az}}$ be the steering vector that denotes beam $k$ in the azimuth plane and  $\mathbf{b}_{\text{pol}}$ be the steering vector that denotes all the beams in the elevation plane. 
For the elevation plane, $\mathbf{b}_{\text{pol}}$ is the vector of length $K$ given by the fixed polar angle $\theta$ and $d_e$ as 
\begin{equation}
    \mathbf{b}_{\text{pol}} = \frac{1}{\sqrt{K}}\left[1, e^{\frac{-j2\pi d_e}{\lambda}\cos\left(\theta\right)},\dotsc , e^{\frac{-j2\pi d_e(K-1)}{\lambda}\cos\left(\theta\right)}\right]^\intercal.
    \label{eq:butler_altitude}
\end{equation}

For the azimuth plane, the signal fed into antenna port $k$ is precoded with the corresponding steering vector, such that the $k$th beam is given by~\cite{Zooghby2005}
\begin{equation}
    \mathbf{b}_{k,\text{az}} = \frac{1}{\sqrt{K}}\left[1, e^{\frac{-j\pi (2k-1)}{K} }, \dotsc , e^{-j\frac{\pi (2k-1)(K-1)}{K}}\right]^\intercal \,.
    \label{eq:butler_azimuth}
\end{equation}
Hence, the steering vector for the $k$th beam with a Butler matrix is calculated as
$\mathbf{b}_{k}=\mathbf{b}_{\text{pol}}\otimes\mathbf{b}_{k,\text{az}}$. On the other hand, the steering vector with digital beamforming and beam steering in the direction of satellite $v$ w.r.t. $u$ is  $\mathbf{b}_{s}^{(u,v)} = \mathbf{a}_t^{(u,v)}/K$. 

\begin{figure}
    \centering
    \includegraphics{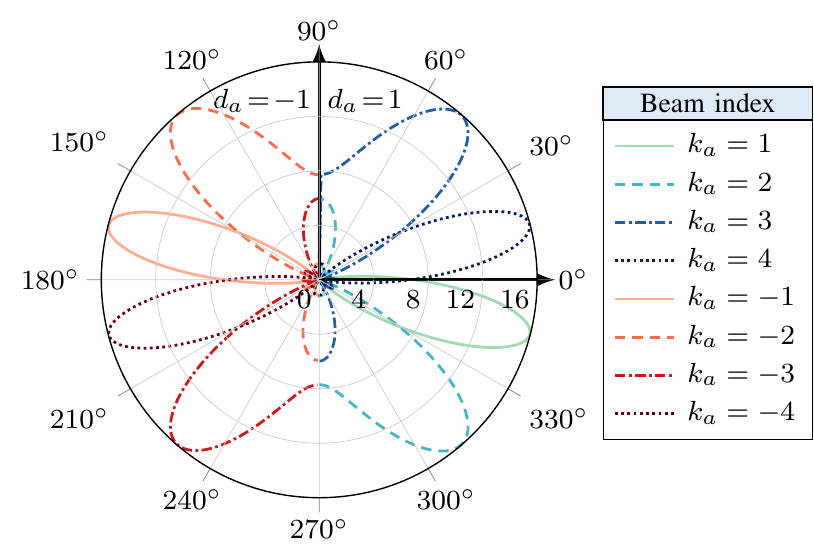}
    \caption{Antenna gains per beam along the azimuth plane with  two $4\times 4$ Butler matrix arrays with $d_e=c/2f$.}
    \label{fig:butler_radiation_pattern}
    \vspace{-2ex}
\end{figure}

 We assume that the satellites possess sufficient shielding so that the gain of an antenna in direction $d_a$ is $0$ in the opposite direction $-d_a$; hence, no power is radiated in direction $-d_a$. Thus, the gain from $u$ to $v$ with beam $k_a^{(u)}$ is given as
\begin{equation}
    g\left(\mathbf{r}^{(u,v)}_t, k_a^{(u)}\right)=\begin{cases}
    |\mathbf{b}_k^H\mathbf{a}^{(u,v)}_t|^2 & \text{if } \cos\left(\phi_t^{(u,v)}\right)d_a>0\\
    0 & \text{otherwise}.
    \end{cases}
    \label{eq:gains}
\end{equation}
Hence, both the fixed beam steering with Butler matrix and the digital beam steering lead to a maximum antenna gain of $10\log(K^2)$\,dBi. Fig~\ref{fig:butler_radiation_pattern} illustrates the radiation pattern of a $4\times4$ antenna array fed by a Butler matrix.

Then, the total antenna gain between satellites $u$ and $v$ with antenna ports $k_a^{(u)}$ and $k_a^{(v)}$, respectively, at time $t$ is given as
\begin{IEEEeqnarray}{rCl}
    G_t\left(u,k_a^{(u)},v,k_a^{(v)}\right) &=& g\left(\mathbf{r}^{(u,v)}_t, k_a^{(u)}\right) g\left(\mathbf{r}^{(v,u)}_t,k_a^{(v)}\right)\IEEEeqnarraynumspace
\end{IEEEeqnarray}

Next, let $m_u(k_a,t)\in\{0,1\}$  be an indicator variable that takes the value of $1$ if satellite $u$ selects antenna beam $k_a$ (i.e., beam $k$ of antenna $d_a$) for communication at time $t$ and $0$ otherwise. 
Thus, the signal-to-noise ratio (SNR) from satellite $u$ to satellite $v\neq u$ at time $t$ is given as
\begin{IEEEeqnarray}{l}
    \text{SNR}\left(u,k_a^{(u)},v,k_a^{(v)},t\right)\IEEEnonumber\\
    \IEEEeqnarraymulticol{1}{r}{\qquad=  \frac{P_tG_t\left(u,k_a^{(u)},v,k_a^{(v)}\right)m_u\left(k_a^{(u)},t\right)m_v\left(k_a^{(v)},t\right)
    }{Bk_\mathrm{B} T_N L(u,v,t)}}\IEEEeqnarraynumspace
\end{IEEEeqnarray}
where $k_\mathrm{B}$ is the Boltzmann constant, $T_N$ is the equivalent noise temperature in Kelvin, and $B$ is the channel bandwidth in Hertz.

The matching occurs periodically with period $\Delta t$ seconds. At each realization, the satellites select their pairs, switch or steer the beam, and select the data rates in the ISL, which remain fixed until the next matching is executed. 

Let $m_{uv}(t)\in\{0,1\}$ be the matching indicator variable that takes the value of $1$ if a link between satellites $u$ and $v$ is established at time $t$. That is, if there is a valid matching at time $t$ that includes satellites $u$ and $v$. Furthermore, to simplify notation, we define the matching indicator
\begin{equation}
    m_u(d_a,t)=\hspace{-1em}\sum_{v\in\mathcal{V}_u(d_a,t)} \hspace{-1em}m_{uv}(t)=\sum_{k_a=d_a}^{nd_a}m_u(k_a,t),
\end{equation}
which takes the value of $1$ if the antenna array of satellite $u$ in direction $d_a$ has established an ISL with another satellite in the corresponding direction $d_a$ at time $t$.

To find appropriate rates for each ISL, we treat the interference as additive-white Gaussian noise (AWGN). Furthermore, we consider the case where the rate for each ISL between $u$ and $v$ is selected as the maximum data rate that can be selected for reliable communication at the endpoints of the matching period $\left[t,t+\Delta t\right]$. Hence, rate selection for each potential ISL is performed based on the  signal-to-interference plus noise ratio (SINR) at $t$ and $t+\Delta t$ by taking the upper bound for the interference. Specifically, the upper bound SNR of the interference at time $t$ at antenna port $k_a^{(v)}$ of the receiver $v$ when $u$ transmits with antenna port $k_a^{(u)}$ and for a specific set of values for the matching variables $\left\{m_{uv}(t)\right\}$ is
\begin{multline}
    I\left(u, k_a^{(u)},v, k_a^{(v)},t\right) \\= \sum_{\substack{i=1}}^{N}\sum_{k_a^{(i)}}\text{SNR}\left(i,k_a^{(i)},v,k_a^{(v)},t\right)\left(1-m_{iv}(t)\right).
    \label{eq:wc_inter}
\end{multline}

Hence, the SINR for a transmission from $u$ to $v$ at time $t$ with antenna ports $k_a^{(u)}$ and $k_a^{(v)}$, respectively, is defined as
\begin{IEEEeqnarray}{c}
    \text{SINR}\left(u,k_a^{(u)},v,k_a^{(v)},t\right)= \frac{\text{SNR}\left(u,k_a^{(u)},v,k_a^{(v)},t\right)}{1+I\left(u,k_a^{(u)},v,k_a^{(v)},t\right)}.\IEEEeqnarraynumspace
    \label{eq:sinr}
\end{IEEEeqnarray}
Finally, the rates for communication are selected as
\begin{IEEEeqnarray}{l}
    R(u,k_a^{(u)},v,k_a^{(v)},t, \Delta t) \IEEEnonumber\\
    \qquad\qquad = B\log_2\Bigg(1 + \min\Big\{\text{SINR}(u,k_a^{(u)},v,k_a^{(v)},t),\IEEEnonumber\\
    \IEEEeqnarraymulticol{1}{r}{\qquad\text{SINR}(u,k_a^{(u)},v,k_a^{(v)},t+\Delta t)}\Big\}\Bigg).
    \label{eq:rates}
\end{IEEEeqnarray}
Note that the latter formulation ensures that outage periods are strictly shorter than the matching period and would only result in zero outage probability if the SINR is concave within the period $\left[t,t+\Delta t\right]$. Instead, zero outage probability can only be guaranteed by selecting the rates based on the minimum SINR over the whole matching period. However, finding the local minima for the SINR is challenging and out of the scope of this paper. If the outage probability must be further reduced, an SINR margin can be defined. 


Now, we define the maximum weighted matching problem as follows.
\begin{IEEEeqnarray}{lCll}
    \maximize\limits_{\substack{\{m_{uv}(t)\}\\\{m_u(k_a,t)\}}}&\quad& \IEEEeqnarraymulticol{2}{l}{\sum_{u=1}^{N}\sum_{\substack{v=1\\v\neq u}}^{N}R(u,k_a^{(u)},v,k_a^{(v)},t, \Delta t))}\IEEEnonumber\\
    &&\IEEEeqnarraymulticol{2}{r}{\times\, m_{uv}(t)\, m_u\left(k_a^{(u)},t\right)\,m_v\left(k_a^{(v)},t\right)}\IEEEeqnarraynumspace\\
    \text{subject to}&& m_u(d_a,t)\in\{0,1\},~ & \forall d_a\in\{-1,1\},u\in\set{V}\IEEEeqnarraynumspace\\
    &&m_{uv}(t)\in\{0,1\},& \forall\, uv\in \set{E}_t, t\\
    &&m_u(k_a,t)\in\{0,1\},& \forall\, u\in \set{V}, k_a, t
    \label{eq:isl_matching_prob}
\end{IEEEeqnarray}

That is, the optimization variables are the satellite pairs and their beams for communication.
Note that the optimal matching is determined by the achievable rate for each ISL and also by its contribution to interference at all the other established ISLs. That is, the interference changes at each step of the matching. Because of this characteristic, our problem is that of a \emph{matching with externalities.} 
Thus, an optimal algorithm to solve the matching must adapt the set of ISLs every time a new ISL is added to the matching and its interference to the other links is calculated. However, this makes the matching problem extremely complex, even though the exact positions of the satellites at any point in time are known. Instead, in the following section we describe a tractable matching algorithm to achieve a near-optimal solution.


\section{Satellite matching algorithm}
\label{sec:algorithms}
In this section, we describe our greedy algorithm to establish the inter-plane ISLs for the case where the satellites communicate an antenna array in each direction $d_a$. 
This algorithm generalizes and extends one of the matching algorithms presented in our previous work~\cite{Leyva-Mayorga2021}. The extensions provided in this section w.r.t. our previous work include the realization of the matching with the data rates calculated from the worst-case SINR at every step of the matching and the mechanism to optimally switch or steer the beams per ISL. Hence, it is applicable to both cases: Butler matrix and beam steering with minor variations described at the end of this section.

The steps to establish the ISLs with a Butler matrix switching network are presented in Algorithm~\ref{alg:matching} and the resulting matching is stored in $\set{M}_t$. Steps 1 to 3 correspond to the initialization of the parameters for the matching at time $t$. Specifically, the matching variables and the interference to all potential links are set to zero at steps 1 and 2. Next, the weights $w_t(uv)$ for each potential satellite pair are defined as the sum of the achievable rates, namely,
\begin{IEEEeqnarray}{l}
    w_t(uv)=\max_{k_a^{(u)},k_a^{(v)}}\Bigg\{\IEEEnonumber\\ R\left(u,k_a^{(u)},v,k_a^{(v)},t, \Delta t\,\middle|\, m_u\left(k_a^{(u)},t\right) m_v\left(k_a^{(v)},t\right)=1\right)\IEEEnonumber\\
    +R\left(v,k_a^{(v)},u,k_a^{(u)},t, \Delta t\,\middle|\, m_u\left(k_a^{(u)},t\right)m_v\!\left(k_a^{(v)},t\right)=1\right)\!\Bigg\}.\IEEEnonumber\\
    \label{eq:weight}
\end{IEEEeqnarray}

At each iteration of the algorithm, the element in $\set{E}_t$ with the greatest weight is selected. Then the satellite pair and antenna beam pair are identified in steps 5 and 6, respectively. Next, the antenna arrays that must be used in the ISL are determined in step 7 and the satellite pairs that are no longer feasible are removed from $\set{E}_t$ in step 8. Step 9 checks if the conditions to establish the ISL are met, that is, if the antenna arrays have no other ongoing connection. If the ISL can be established, the satellites and antenna beams are added to the matching (step 10), and the matching indicator variables are updated (step 11). With the updated matching variables, the upper bound of the contribution to interference by the newly-added ISL is calculated in step 12. Step 13 is optional and allows to update the weights of the remaining feasible ISLs based on the updated interference. This process is performed until $\set{E}_t$ is empty. At the end of the matching, the rates for each ISL are calculated with the upper bound for the interference. 

\begin{algorithm} [t]
	\centering
	\caption{Greedy satellite matching with multiple beams.}
	\begin{algorithmic}[1] 
	\renewcommand{\algorithmicrequire}{\textbf{Input:}}
		\renewcommand{\algorithmicensure}{\textbf{Output:}}
		\REQUIRE Set of feasible weighted edges  $\set{E}_t$ and $\set{E}_{t+\Delta t}$ 
		\REQUIRE Antenna array configuration: $K$, $d_e$, and $\theta$
		\STATE $\set{M}_t=\emptyset$, $\left\{m_{uv}(t), m_u\left(k_a,t\right ), m_u\left(k_a,t\right)\right\} \leftarrow 0$ 
		\STATE $I\left(u,k_a^{(u)},v,k_a^{(v)},t\right)=I\left(u,k_a^{(u)},v,k_a^{(v)},t+\Delta t\right)=0$ for all $u,v, k_a^{(u)},k_a^{(v)}$ 
		\STATE Calculate $w_t(e)$ for all $e\in\set{E}_t$ as in~\eqref{eq:weight}
		\WHILE { $\mathcal{E}_t \neq \emptyset$}
		\STATE $uv\leftarrow e^*:w(e^*)\geq w(e)$ for all $e\in\set{E}_t$
		\STATE $\left(k_a^{(u*)},k_a^{(v*)}\right)\leftarrow \argmax\limits_{\left(k_a^{(u)},k_a^{(v)}\right)}R'\left(u,k_a^{(u)},v,k_a^{(v)}\right)$
		\STATE $d_a^{(u)}=\sgn k_a^{(u*)}$ , $d_a^{(v)}=\sgn k_a^{(v*)}$
		\STATE $\set{E}_t\leftarrow\set{E}_t\setminus\left\{\{uv'\in \set{V}_u(d_a^{(u)},t)\}\cup\{vv'\in \set{V}_v(d_a^{(v)},t)\}\right\}$
		\IF {$m_{u}\left(d_a^{(u)},t\right)==0$ \AND $m_{v}\left(d_a^{(v)},t\right)==0$ }
		\STATE $\set{M}_t\leftarrow \set{M}_t\cup \{u,v,k_a^{(u*)},k_a^{(v*)}\}$
		\STATE $\left\{m_{uv}(t),m_{u}\left(d_a^{(u)},t\right),m_{u}\left(k_a^{(u*)},t\right)\right\}\leftarrow 1$ and
		
		$\left\{m_{v}\left(d_a^{(v)},t\right),m_{v}\left(k_a^{(v*)},t\right)\right\}\leftarrow 1$
		\STATE  Update $I\left(u,k_a^{(u)},v,k_a^{(v)},t\right)$ and $I\left(u,k_a^{(u)},v,k_a^{(v)},t+\Delta t\right)$ for all $\{u,v,k_a^{(u)},k_a^{(v)}\}$ as in~\eqref{eq:wc_inter}
	 \STATE Optional: Update $w_t(e)$ for all $e\in\set{E}_t$ as in~\eqref{eq:weight}
		\ENDIF		
		\ENDWHILE
	\ENSURE $\set{M}_t$
	\end{algorithmic}  
	\label{alg:matching}
\end{algorithm}
The complexity of each iteration of Algorithm~\ref{alg:matching} is determined by the updates of the weights in step 12. There are $4N^2K^2$ possible pairs of antenna ports, hence calculating the contribution to the interference towards each of these has a complexity $\set{O}\left(N^2K^2\right)$. This process must be performed each time a new ISL (i.e., pair of antenna ports) is added to the matching, which occurs $\set{O}(N)$ times throughout the matching. Hence, the complexity of Algorithm~\ref{alg:matching} is $\set{O}\left(N^3K^2\right)$.

For the case with digital beam steering, Algorithm~\ref{alg:matching} can be easily modified as follows.
\begin{enumerate}
    \item During initialization, calculate the optimal beam steering vector $\mathbf{b}_{d_a,\text{opt}}^{(u,v)}$ for all $uv\in\set{E}_t$.
    \item Calculate the gains for all $G_t(u,d_a^{(u)},v,d_a^{(v)})$ by substituting $\mathbf{b}_{k}$ with $\mathbf{b}_{d_a,\text{opt}}^{(u,v)}$ in~\eqref{eq:gains}.
    \item Perform the matching as indicated by Algorithm~\ref{alg:matching} by setting $k_a^{(u)}=d_a$  and selecting the appropriate steering vectors $\mathbf{b}_{d_a,\text{opt}}^{(u,v)}$ and $\mathbf{b}_{d_a,\text{opt}}^{(v,u)}$ at step 7.
    \item Skip the optional step 13. 
\end{enumerate}
With these modifications, the complexity of the matching algorithm for the beam steering case is reduced to $\set{O}\left(N^3\right)$.

Two benchmarks were defined to assess the performance of the use of antenna arrays with Butler matrix or beam steering. The first one is the case with a half-wave dipole antenna in each direction $d_a$, inclined so the maximum radiation is pointed towards $\theta$. Algorithm~\ref{alg:matching} is directly applied for the matching. The second benchmark is the case with a similar $K\times K$ antenna array in a simplified interference-free scenario.

The following key performance indicators have been defined. The main performance indicator is the average sum of data rates per matching as given by \eqref{eq:isl_matching_prob}. To evaluate the impact of interference, we calculate the relative loss in the sum of data rates w.r.t. the simplified scenario with no interference. 

Establishing the cross-seam ISLs -- those between satellites in orbital planes $1$ and $7$ -- presents a major challenge, not only for the pointing of the antennas but also due to the large Doppler shift~\cite{Su2019}. Hence, these ISLs are disabled by default but we also evaluate the increase in the sum of rates when cross-seam ISLs are enabled.




\section{Results}
\label{sec:results}

The parameters selected for performance evaluation are listed in Table~\ref{tab:sim_params}. A Walker star constellation is considered. The selected orbital separation of $4$~km leads to $5$ seconds or around $0.086$\% of difference between the periods of neighboring orbital planes. During the performance evaluation, it was observed that $98$\% of the inter-plane ISLs were established with satellites with a relative polar angle $100^\circ\pm 10^\circ$. Hence, the fixed polar angle for the Butler matrix and for the half-wave dipole was set to $\theta=100^\circ$.

The results were obtained by simulation in Python. At each simulation instance (i.e., realization of the matching) the position of the satellites, the antenna radiation pattern, the achievable data rates, and the interference are calculated and the matching is performed as described in Algorithm~\ref{alg:matching}. To obtain the results, at least $500$ simulations instances per configuration were run: $10$ different initial placements of the satellites were considered to account for the displacement of the orbital planes through time due to the orbital separation. Hence, at least $100$ simulations were run at consecutive intervals of $\Delta t$ seconds after each of the $10$ initial placements. 

In our experiments, the loss in the sum of rates due to interference with the Butler matrix is as low as $0.0068$\% with $K=1$ and as high as $0.0306$\% with $K=8$. Furthermore, the difference between performing or skipping step 13 of Algorithm~\ref{alg:matching} was lower than $0.0008$\% and no conclusive results were obtained on which version of Algorithm~\ref{alg:matching} results in higher rates. For digital beam steering, the loss due to interference is concave with a maximum of $0.6115$\% at $K=8$.

\begin{table}[t]
    \centering
     \caption{Parameter settings for performance evaluation}
    \begin{tabular}{@{}lll@{}}
    \toprule
    \multicolumn{1}{@{}l}{Parameter} & Symbol & Setting\\\midrule
    Number of orbital planes & $P$ & $7$\\
    Number of satellites & $N$ & $140$\\
    Inclination of the orbital planes & $\delta$ & $98.6^\circ$\\
    Minimum altitude of deployment & $h_1$ & $600$\,km\\
    Orbital separation & $\Delta h$ & $4$\,km\\
    Transmission power & $P_t$ & $10$\,W\\
    System bandwidth & $B$ & $200$\,MHz\\
    Carrier frequency & $f$ & $20$\,GHz\\
    Noise temperature & $T_N$ & $324.81$\,K\\
     Matching period & $\Delta t$ & $\{10,30,60\}$\,s\\
    Number of antenna ports/beams & $K$ & $\{1,2,4,8\}$\\
    Polar angle for Butler matrix and dipole& $\theta$ & $100^\circ$\\
    Distance between antenna elements & $d_e$ & $c/2f$\\\bottomrule
    \end{tabular}
   
    \label{tab:sim_params}
\end{table}

\begin{table}[t]
    \centering
    \caption{Average sum of rates with different antenna configurations}
    \renewcommand{\arraystretch}{1.1}
    \begin{tabular}{@{}rlrrr@{}}
    \toprule
         & \phantom{a}& \multicolumn{3}{c}{Average sum of rates (Mbps)}\\\cmidrule{3-5}
     & & \multicolumn{1}{c}{$\Delta t=10$} & \multicolumn{1}{c}{$\Delta t=30$} & \multicolumn{1}{c}{$\Delta t=60$}\\\midrule
    \multicolumn{1}{@{}l}{\textbf{Isotropic ($K=1$)}}& & $1.23527$ & $0.7568$ & $0.5135$ \\
    \multicolumn{1}{@{}l}{\textbf{Dipole}} & & $3.1164$ & $1.8727$& $1.3345$\\
    \multicolumn{1}{@{}l}{\textbf{Butler matrix}} & & & & \\
     $K=2$ & &$10.9524$ & $7.2658$ & $5.1212$\\
     $K=4$ & &$102.4879$ & $72.3870$ & $53.5573$\\
     $K=8$ & & $706.6720$ & $601.5892$ & $477.9366$\\
    \multicolumn{1}{@{}l}{\textbf{Beam steering}} & & & & \\
     $K=2$ & & $18.7919$ & $10.9284$ & $7.7027$\\
     $K=4$ & & $265.0107$ & $154.5756$ & $107.9056$\\
     $K=8$ & &$3073.3669$ & $1943.3217$ & $1347.3270$\\
         \bottomrule
    \end{tabular}
    \label{tab:results}
\end{table}
As a starting point, Table~\ref{tab:results} shows the average sum of rates with different numbers of antenna ports $K$ and  matching periods $\Delta t$. The results with the half-wave dipole antenna and with an isotropic antenna (i.e., $K=1$) are also included.

It is clear from Table~\ref{tab:results} that the sum of rates increases rapidly with $K$ and that any $K>1$ outperforms case with the dipole antenna. In particular, with the Butler matrix, doubling the number of antennas increases the rates by nearly one order of magnitude and an even greater increase is observed with digital beam steering. The reason for this is that, as the beams become narrower, the pointing precision becomes more important and digital beam steering is capable of precisely pointing the beam to achieve the greatest gain in the links. As a consequence, the benefits of digital beam steering w.r.t. beam switching with the Butler matrix increase with $K$ but so does the complexity of the matching. However, the matching algorithm does not increase the computational load in the satellites, as all the settings can be computed in a centralized manner and simply distributed throughout the constellation.

Table~\ref{tab:results} also shows that the sum of rates increases rapidly by decreasing the matching period. Specifically, the sum of rates with $\Delta t =10$\,s is approximately twice the sum of rates with $\Delta t =60$\,s in most cases. This is because the highest rates are achieved near the crossing points of the orbits, where the relative positions change rapidly and where shorter matching periods allow to select higher rates. Despite its benefits, $\Delta t$ cannot be reduced arbitrarily as frequent link switching may cause problems, for instance, when implementing a routing algorithm. Furthermore, establishing an ISL requires, at least, a handshake between the involved satellites, whose time to complete lies in the order of $20$\,ms round-trip due to the propagation delay between the satellites in the considered constellation. Hence, the control overhead increases as $\Delta t$ decreases and a lower limit for $\Delta t$ must be established. 

\begin{figure}
    \centering
    \includegraphics{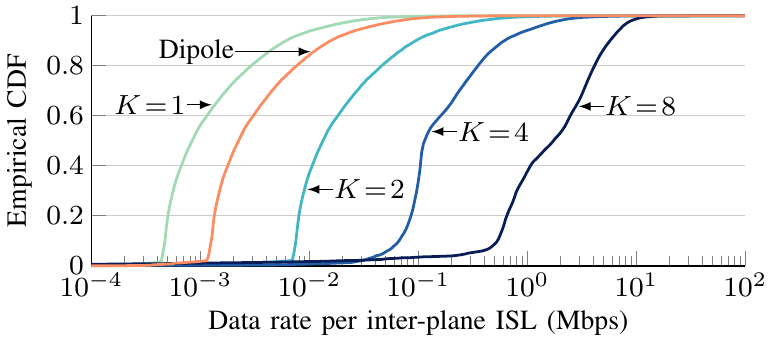}
    \caption{CDF of the selected data rates per ISL with Butler matrix having $K=\{1,2,4,8\}$ antenna ports and a half-wave dipole antenna for $\Delta t =30$\,s.}
    \label{fig:rates_beams}
    \vspace{-2ex}
\end{figure}

Next, Fig.~\ref{fig:rates_beams} shows the CDF of the rates per ISL for the Butler matrix with $K=\{1,2,4,8\}$ and for the half-wave dipole antenna for $\Delta t=30$\,s. Clearly, the rates per ISL increase with $K$ in a similar proportion as the sum of rates. Furthermore, with $K=8$, less than $5$\% of the ISLs achieve a rate lower than $300$\,kbps and less than $5$\% achieve a rate higher than $7.5$\,Mbps. In contrast, the difference between the highest and lowest rates with $K=1$ is much greater. Finally, the rates with the half-wave dipole antenna are, as expected, only slightly higher than those for $K=1$.

\begin{figure}
    \centering
    \includegraphics{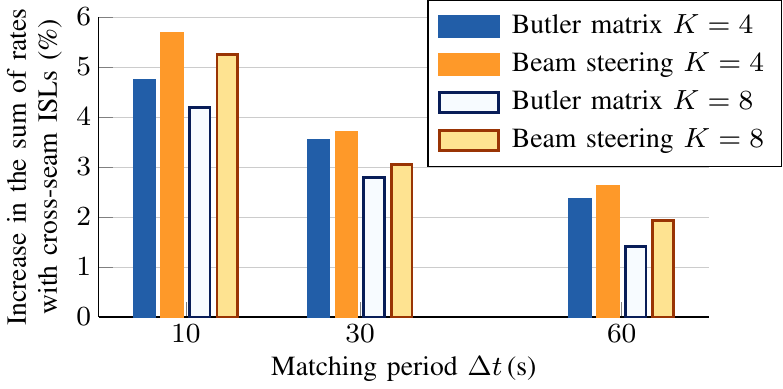}
    \caption{Sum rate increase due to enabling cross-seam ISLs for the Butler matrix and digital beam steering with $K=\{4,8\}$ and for $\Delta t=\{10,30,60\}$.}
    \label{fig:increase_cs}
    \vspace{-2ex}
\end{figure}

Lastly, we investigate the effect of enabling the cross-seam ISLs in the sum of rates. Here, the potential effect of the Doppler shift is neglected to focus on beam steering. Fig.~\ref{fig:increase_cs} shows the relative increase in the sum of rates for $K=\{4,8\}$ and $\Delta t=\{10,30,60\}\,s$ with enabled cross-seam ISLs w.r.t. to the baseline scenario where these are disabled. Note that 1) the gains of enabling the cross-seams ISLs increase as $\Delta t$ decreases and 2) the gains with $K=4$ are greater than with $K=8$. The reason for these is that these satellites in cross-seam ISLs are moving in nearly different directions and reach relative velocities of up to $12$\,km/s. Consequently 1) the SINR in the links changes rapidly and shorter matching periods allow to select higher rates and 2) wider beams are better suited to connect satellites with such high relative velocities. Specifically, the sum of rates with $K=4$ increased $4.7$\% and $5.6$\% with $\Delta t=10$\,s for the Butler matrix and for digital beam steering, respectively. On the other hand, the maximum gain with $K=8$ is $5.2$\%. Finally, the gains are, in most cases, greater with the Butler matrix than with digital beam steering, as the latter allows for an optimal placement of the beams. Nevertheless, an increase in the sum of rates of $5$\% is modest and may be further reduced due to Doppler shift.


\section{Conclusion}
In this paper, we investigated the benefits of antenna arrays with either beam switching with Butler matrix or digital beam steering for the inter-plane ISLs in LEO satellite constellations. Furthermore, we presented a matching algorithm that: 1) takes into account interference; 2) is applicable to cases where numerous beams and/or antenna ports are available per antenna; and 3) does not increase the computational load in the satellites.
Our results show how increasing the number of antenna elements and decreasing the matching period increases performance. Furthermore, we observed that the cross-seam ISLs can only be established efficiently by defining a sufficiently short matching period. However, the increase in the sum of rates by establishing cross-seam ISLs is only $5$\%. Finally, we observed that the impact of interference on the performance in the selected scenario is minimal. 

\bibliographystyle{IEEEtran}
\bibliography{IEEEabrv,bib}
\end{document}